\documentclass[11pt]{article}
\usepackage{moriond,epsfig}

\def\be{\begin{equation}}
\def\ee{\end{equation}}
\def\bea{\begin{eqnarray}}
\def\eea{\end{eqnarray}}

\newcommand{\ba}{\begin{eqnarray}}
\newcommand{\ea}{\end{eqnarray}}
\newcommand{\dis}{\displaystyle}

\newcommand{\im}{\mbox{Im}\,}

\begin{document}
\vspace*{4cm}
\title{DETERMINATION OF $|V_{us}|$ FROM HADRONIC $\tau$ DECAYS}

\author{ Elvira G\'amiz$^a$, 
Matthias Jamin$^b$,
Antonio Pich$^c$,
 Joaquim Prades$^d$ 
and  Felix Schwab$^{e, f}$}

\address{$^a$ Department of Physics $\&$ Astronomy,
University of Glasgow\\ Glasgow G12 8QQ, United Kingdom.\\
$^b$ ICREA and IFAE, Universitat Aut\`onoma de Barcelona\\
E-08193 Bellaterra, Barcelona, Spain.\\
$^c$ Departament de F\'{\i}sica Te\`orica, IFIC, 
Universitat de Val\`encia--CSIC\\ Apt. de Correus 22085, E-46071
Val\`encia, Spain.\\
$^d$ Centro Andaluz de F\'{\i}sica de las Part\'{\i}culas
Elementales (CAFPE) and Departamento de
 F\'{\i}sica Te\'orica y del Cosmos, Universidad de Granada \\
Campus de Fuente Nueva, E-18002 Granada, Spain.\\
$^e$  Physik Department, Technische
Universit\"at M\"unchen\\ D-85747 Garching, Germany.\\
$^f$ Max-Planck-Institut f\"ur Physik --
Werner-Heisenberg-Institut\\ D-80805 M\"unchen, Germany.}

\maketitle\abstracts{
      The recent update of the strange spectral function and the moments of 
   the invariant mass distribution by the OPAL collaboration from hadronic 
   $\tau$ decay data are employed to determine $|V_{us}|$ as well as $m_s$. Our 
   result, $|V_{us}|=0.2208\pm0.0034$, is competitive to the standard 
extraction 
   of $|V_{us}|$ from $K_{e3}$ decays and to the new proposals 
to determine it. 
   Furthermore, the error associated to our determination of $|V_{us}|$ can be 
   reduced in the future since it is dominated by the experimental 
   uncertainty that will be eventually much improved by the B-factories 
   hadronic $\tau$ data. 
   Another improvement that can be performed is the simultaneous fit of 
   both $|V_{us}|$ and $m_s$ to a set of moments of the hadronic $\tau$ decays 
   invariant mass distribution, which will provide even a more accurate 
   determination of both parameters.}

\section{Introduction}

Already in the past, hadronic $\tau$ decays have served as an interesting 
source to study low energy QCD under rather clean conditions~\cite{BNP92} 
and obtain information on parameters of the Standard Model, 
like the strong coupling~\cite{ALEPH98,OPAL99}, the strange quark mass 
or some non-perturbative condensates. The 
determination of $\alpha_s$, for example, has been performed using 
$\tau$ decay data with a precision competitive to the current world average.

At present, the hadronic $\tau$ decay width 
\be
\label{Rtau}
R_\tau \equiv 
\frac{\Gamma\left[\tau^- \to {\rm hadrons} (\gamma)\right]}
{\Gamma\left[ \tau^- \to e^- \overline{\nu}_e \nu_\tau (\gamma)\right]}
\ee
as well as invariant mass distributions, have reached a high precision status 
thanks to the data of the LEP  experiments ALEPH~\cite{ALEPH98}
and OPAL~\cite{OPAL99}  at CERN
 and the CESR experiment CLEO ~\cite{CLEO93} at Cornell.

In particular, the experimental measurements of the strange spectral function 
~\cite{ALEPH99,CLEO03,OPAL04} 
have allowed the analysis of the SU(3) breaking corrections in the 
semi-inclusive $\tau$ decay width into Cabibbo suppressed modes with strange 
particles, providing a way to compute the strange quark mass $m_s(M_{\tau})$ 
~\cite{CHD97,PP98,CKP98,PP99,KM00,CDGHPP01,GJPPS03,GJPPS04a}. Recently, we 
have pointed out that this approach to obtain the strange mass from the 
hadronic $\tau$ decays depends sensitively on the modulus of the 
Cabibbo--Kobayashi--Maskawa matrix element $|V_{us}|$ \cite{GJPPS03}. 
It appears then natural to turn things around and, with an 
input for $m_s$ obtained from other sources, to actually determine $|V_{us}|$. 
The great advantage of the determination of this CKM matrix element from 
$\tau$ decays in comparison with other calculations is that the experimental 
uncertainty, which as we will see is the main source of error in the 
calculation, is expected to be reduced drastically at the present B-factories 
BABAR and BELLE.
        
\section{Theoretical framework}

The main quantity of interest for the following analysis is the hadronic 
decay rate of the $\tau$ lepton defined in (\ref{Rtau}). The basic objects 
one needs to perform the QCD analysis of $R_{\tau}$ and related observables 
are Green's two-point functions for vector 
$V^{\mu}_{ij} \equiv \overline{q}_i \gamma^\mu q_j$ and axial-vector
$A^{\mu}_{ij} \equiv \overline{q}_i \gamma^\mu \gamma_5 q_j$ 
color singlets, 
\ba
\Pi^{\mu\nu}_{V,ij}(q)\equiv i \int {\rm d}^4 x \, e^{i q\cdot x}
\langle 0 | T \left([V^\mu_{ij}]^\dagger(x) V^\nu_{ij}(0)\right)
| 0 \rangle \, , \nonumber \\
\Pi^{\mu\nu}_{A,ij}(q)\equiv i \int {\rm d}^4 x \, e^{i q\cdot x}
\langle 0 | T \left([A^\mu_{ij}]^\dagger(x) A^\nu_{ij}(0)\right)
| 0 \rangle \, .
\ea
 The subscripts $i,j$ denote light  quark flavors (up, down
and strange). These correlators admit the Lorentz decompositions
\ba
\Pi^{\mu\nu}_{ij,V/A}(q) &=&
\left(-g_{\mu\nu} q^2 + q^\mu q^\nu \right) \, \Pi^T_{ij,V/A}(q^2)
+q^\mu q^\nu \, \Pi^L_{ij,V/A}(q^2)\, 
\ea
where the superscripts  in the transverse and longitudinal
components denote the spin $J=1 \, (T)$ and $J=0 \, (L)$
in the hadronic rest frame. Theoretically, $R_{\tau}$ can 
be expressed as an integral of the imaginary part of these correlators 
over the invariant mass $s=p^2$ of the final state hadrons~\cite{BNP92}
\be
\label{RTauth}
R_\tau = 12\pi\!\!\int\limits_0^{M_\tau^2}\!\frac{ds}{M_\tau^2}\,\biggl(
1-\frac{s}{M_\tau^2}\biggr)^2 \biggl[\biggl(1+\frac{2s}{M_\tau^2}\biggr)
\im\Pi^T(s)+\im
\Pi^L(s)\biggr], 
\ee
where the appropriate combinations of two-point correlation functions are
\begin{eqnarray}
\Pi^J(s) \equiv |V_{ud}|^2\,\Big[\,\Pi^{V,J}_{ud}(s) +
\Pi^{A,J}_{ud}(s)\,\Big] 
+ |V_{us}|^2\,\Big[\,\Pi^{V,J}_{us}(s) + \Pi^{A,J}_{us}(s)\,\Big]\,,
\end{eqnarray}
with $V_{ij}$ being the corresponding matrix elements of the CKM matrix. 

Experimentally, one can disentangle vector from axialvector contributions 
in the Cabibbo-allowed ($\bar ud$) sector, whereas such a separation is 
problematic in the Cabibbo-suppressed ($\bar us$) sector. We can then 
decompose $R_{\tau}$ both experimentally and theoretically into
\be
R_\tau \equiv R_{\tau,V} + R_{\tau,A} + R_{\tau,S}. 
\ee

Additional information can be inferred from the measured invariant mass 
distribution of the final state hadrons, through the analysis of the moments 
\be
\label{OPE}
R_\tau^{(k,l)} \equiv {\dis \int^{M_\tau^2}_0}
{\rm d} s \left(1-\frac{s}{M_\tau^2}\right)^k \, 
\left( \frac{s}{M_\tau^2} \right)^l \, \frac{{\rm d} R_\tau}{{\rm d} s}, 
\ee
that can be calculated in analogy to the $\tau$ decay rate ($R_{\tau}=
R_{\tau}^{(0,0)})$ and, in particular, can be also decomposed  
into Cabibbo-allowed and Cabibbo suppressed contributions.

The theoretical study of $R_{\tau}$ and its moments is based on the Operator 
Product Expansion (OPE) of the relevant correlators. In this framework, the 
moments $R_{\tau}^{(k,l)}$ can be written as
\ba\label{momentOPE}
R_\tau^{(k,l)} \equiv N_c S_{\rm EW} 
\Big\{ (|V_{ud}|^2 + |V_{us}|^2) \,  \left[ 1 + \delta^{(k,l)(0)}\right]
 + {\dis \sum_{D\geq2}} \left[ |V_{ud}|^2 \delta^{(k,l)(D)}_{ud}
+ |V_{us}|^2 \delta^{(k,l)(D)}_{us} \right] \Big\} \, . 
\ea
The electroweak radiative correction 
$S_{\rm EW}=1.0201\pm0.0003$~\cite{Sew} has been pulled out explicitly
and $\delta^{(k,l)(0)}$ denotes  the purely perturbative
dimension-zero contribution, that is the same for the Cabibbo-allowed and 
Cabibbo-suppressed contributions.  The symbols $\delta^{(k,l)(D)}_{ij}$
stand for  higher dimensional corrections in the OPE from 
dimension  $D\geq 2$ operators which contain  implicit $1/M_\tau^D$
suppression factors~\cite{BNP92,PP98,PP99,CK93}. The most important 
corrections are the dimension $D=2$ proportional to $m_s^2$ and the dimension 
$D=4$ proportional to $m_s\langle \bar qq\rangle$. 

The separate measurement of strange and non-strange contributions to the decay 
width of the $\tau$ lepton~\cite{ALEPH99,CLEO03,OPAL04} allows one to pin down the 
flavour SU(3)-breaking effects, dominantly induced by the strange quark mass,  
through the differences
\ba
\label{deltaR}
\delta R^{(k,l)}_\tau \!\!&\equiv\!\!& 
\frac{R^{(k,l)}_{\tau,V+A}}{|V_{ud}|^2}-
\frac{R^{(k,l)}_{\tau,S}}{|V_{us}|^2} = 
N_c \, S_{EW} \, {\dis \sum_{D\geq 2}}
 \left[ \delta^{(k,l)(D)}_{ud}
-\delta^{(k,l)(D)}_{us}\right] \, . 
\ea
Many theoretical uncertainties drop out in these observables since they 
vanish in the SU(3) limit.

\section{Calculation of $|V_{us}|$}

The large sensitivity of the SU(3) breaking quantities 
$\delta R_{\tau}^{(k,l)}$ allows us to obtain a determination of the CKM 
matrix element $|V_{us}|$ using as input a fixed value of $m_s$. Since the 
sensitivity to $|V_{us}|$ is strongest for the moment with $k=l=0$, where also 
the theoretical uncertainties are smallest, we used this moment in our 
calculation. From (\ref{deltaR}),
\be\label{vusexp}
|V_{us}|^2 = 
\frac{R^{(0,0)}_{\tau,S}}
{\frac{\dis R^{(0,0)}_{\tau,V+A}}
{\dis |V_{ud}|^2}-\delta R^{(0,0)}_{\tau,{\rm th}}} \, .
\ee

In the OPE of $\delta R^{(0,0)}_{\tau,{\rm th}}$ in (\ref{deltaR}) we include 
the dimension two corrections  $\delta^{(k,l)(2)}_{ij}$ that are known 
at ${\cal O}(a^3)$ for $J=L$ component and at ${\cal O}(a^2)$
for $J=L+T$ component --see~\cite{PP98,PP99} for references. The 
${\cal O}(a^3)$ $J=L+T$ are also known~\cite{CHE04} but 
the results obtained in~\cite{GJPPS04a} and discussed here do not contain these
corrections. We leave a full analysis which take them into account
for a future publication~\cite{GJPPS04b}. Dimension four corrections 
$\delta^{(k,l)(4)}_{ij}$ are fully included while the
dimension six corrections $\delta^{(k,l)(6)}_{ij}$ were
estimated to be of the order or smaller than the error
of the dimension four ~\cite{PP99}.

An extensive analysis of the perturbative series
for the dimension two corrections was done in
~\cite{PP98}. The conclusions there were that  while the 
perturbative series for $J=L+T$ converges very well
the one for the $J=L$ behaves very badly, adding important uncertainties 
to the theoretical calculation of $\delta R_{\tau}^{(k,l)}$. A natural 
remedy to solve this 
problem is to replace the QCD expressions of scalar and pseudoscalar 
correlators by corresponding phenomenological hadronic parametrizations,  
much more precise than their QCD counterpart~\cite{GJPPS03} due to the fact 
that it is dominated by far by the well known kaon pole --see~\cite{GJPPS03} 
for details. 
\begin{table}[htb]
\label{tab1}
\caption{Comparison between the OPE and the phenomenological
hadronic parametrizations explained in the text
 for the longitudinal component of $R^{(0,0)}_{\tau,V/A}$.}
\begin{center}
\begin{tabular}{cccc}
\hline 
& $R^{(0,0)L}_{us,A}$ & $R^{(0,0)L}_{us,V}$ & 
$R^{(0,0)L}_{ud,A}\times 10^{3}$ \\
\hline 
OPE & $-0.144\pm0.024$ &$-0.028\pm0.021$ & $-7.79\pm0.14$ \\
Pheno. & $-0.135\pm0.003$ &$-0.028\pm0.004$ & $-7.77\pm0.08$ \\
\hline
\end{tabular}
\end{center}
\end{table}
The results we obtained within the QCD correlators and within the 
phenomenological hadronic parametrized correlators are showed in Table 1. 
The longitudinal contributions calculated with the two different 
descriptions of the spectral functions are very similar, but the errors are 
much lower using the phenomenological parametrization. Therefore, using 
phenomenology to describe the J=L component of $\delta R_{\tau}^{(k,l)}$ 
reduces considerably the theoretical uncertainty.

We use as input value $m_s( 2 \,{\rm GeV})=(95\pm20)$ MeV
which includes the most recent determinations of $m_s$
from QCD Sum Rules~\cite{MK02,JOP02,NAR99}, 
lattice QCD~\cite{HAS04} and $\tau$ hadronic data
~\cite{CHD97,PP98,CKP98,PP99,KM00,CDGHPP01,GJPPS03}.
With this strange quark mass input and taking $\delta R_{\tau}^{(0,0)L+T}$ 
and $\delta R_{\tau}^{(0,0)L}$ from the QCD OPE and phenomenology 
respectively, as explained above, one can calculate 
$\delta R^{(k,l)}_\tau$ in (\ref{deltaR}) from theory
\ba
\delta R^{(0,0)}_{\tau,{\rm th}}
= (0.162\pm0.013) + (6.1 \pm 0.6) \, m_s^2 
-(7.8\pm0.8) \, m_s^4 = 0.218 \pm 0.026,
\ea
where   $m_s$ denotes the strange quark mass in MeV units,
defined  in the $\overline{MS}$ scheme  at 2 GeV.

In order to obtain a value of $|V_{us}|$ from (\ref{vusexp}), we also need the 
experimentally measured Cabibbo-allowed $R^{(0,0)}_{\tau,V+A}$ 
and Cabibbo-suppressed $R^{(0,0)}_{\tau,S}$ contributions to the $\tau$ decay rate. 
OPAL has recently updated the strange spectral function
in~\cite{OPAL04}. In particular, they measure
a larger branching fraction $B(\tau^-\to K^- \pi^+ \pi^- \nu)$
which agrees with the previously one measured by CLEO~\cite{CLEO03}. 
From the OPAL data and using (\ref{vusexp}) we get the result
\ba
\label{vus}
|V_{us}|&=&0.2208\pm0.0033_{\rm exp}\pm 0.0009_{\rm th}
= 0.2208 \pm 0.0034 \, ; 
\ea
where we have used as input the PDG value for $|V_{ud}|=0.9738\pm0.0005$. 
The most important feature of this determination is the small theoretical 
error, that leaves the final uncertainty dominated by the experimental input. 
One can expect then that the better data sets from BABAR and BELLE will 
reduce significantly the error of our determination by improving the experimental 
input. Nevertheless, already now, 
our result is competitive to the standard extraction of $|V_{us}|$ from 
semileptonic kaon 
decays~\cite{lr:84,e865:03,ktev:04,ckm:03,JOP04,Becirevic04,BT03,CNP04} 
and a new determination from $f_K/f_{\pi}$ as extracted from 
lattice~\cite{Marciano,MILC04}.

One can use the value of $|V_{us}|$ thus obtained in (\ref{vus}) and determines 
the strange quark mass from higher moments. The weighted average of $m_s$ 
calculated for the different moments gives
\ba
\label{ms}
m_s(M_\tau)= 84 \pm 23 \, {\rm MeV} \, \quad\quad
\Rightarrow  \quad\quad m_s(2\,{\rm GeV})= 81 \pm 22 \, {\rm MeV} \ .
\ea
For details about the results for the different moments and the sources of 
the uncertainties see~\cite{GJPPS04a}. 

In our previous analysis based on the ALEPH data~\cite{GJPPS03}, it was observed 
that $m_s$ displayed a strong dependence on the number of the moment $k$, 
decreasing with increasing $k$. With the recent CLEO and OPAL results finding 
a larger branching fraction  $B(\tau^-\to K^- \pi^+ \pi^- \nu)$, this dependence 
is much reduced, although still visible. This issue needs to be clarified with 
the help of better experimental data.

\section{Simultaneous fit of $m_s$ and $|V_{us}|$}

The ultimate procedure to determine both $|V_{us}|$ and $m_s$ from $\tau$ 
hadronic decays, will be a simultaneous 
fit of both to a certain set of $(k,l)$ moments.  A detailed
study  including theoretical and experimental correlations
will be presented elsewhere~\cite{GJPPS04b}. 

In~\cite{GJPPS04a}, we restrict ourselves to a simplified approach where all 
correlations were neglected. For this simultaneous fit of $|V_{us}|$ and $m_s$ 
we use the five OPAL moments~\cite{OPAL04} from 
$R^{(0,0)}_\tau$ to $R^{(4,0)}_\tau$. The central values we find from this exercise 
are
\be
|V_{us}|=0.2196 \hspace{0.5cm} {\rm and} \hspace*{0.5cm} 
m_s(2 \,{\rm GeV}) = 76 \, {\rm MeV} \,.
\ee
These values are in very good agreement with our previous results in (\ref{vus}) 
and (\ref{ms}). We expect that the uncertainties on these results will be smaller 
than the individual errors in (\ref{vus}) and (\ref{ms}), but only  slightly 
since the correlations between different moments are rather strong.

\section{Conclusions and remarks}

Using the strange spectral function updated by OPAL~\cite{OPAL04}, we get
\be
|V_{us}|=0.2208\pm0.0033_{\rm exp}\pm 0.0009_{\rm th} \, , 
\ee
and 
\be
\label{ms2}
m_s(2 \,{\rm GeV}) = 81\pm 22 \, {\rm MeV} \, .
\ee 
This result is expected to be highly improved in the near future due to the 
fact that the error is dominated by the experimental uncertainty and that 
uncertainty can be reduced with better data samples from BABAR and BELLE. 
But already now, the high precision $\tau$ decay data from  
ALEPH and OPAL at LEP and CLEO at CESR provide competitive results for $|V_{us}|$
and $m_s$. The combined fit to determine both quantities including 
theoretical and experimental correlations is underway.

The actual status of the $|V_{us}|$ determinations
 has been nicely reviewed recently  in~\cite{CMS04,ROS04}.
 Though the CKM unitarity  discrepancy has certainly
decreased with the new theoretical and experimental
advances, the situation is not yet as good as one could
wish, and an accurate determination for $|V_{us}|$ with the eventual precise  
measurement of the strange spectral function at BABAR and BELLE is desirable. 
With the value of $|V_{us}|$ in (\ref{vus}) and using the PDG value 
$|V_{ud}|=0.9738\pm0.0005$, one finds
\be
1-|V_{ud}|^2-|V_{us}|^2-|V_{ub}|^2=(2.9\pm1.8)\cdot 10^{-3},
\ee
so that unitarity is violated only at the $1.6\,\sigma$ level.

There are some open questions that will have also to be addressed~\cite{GJPPS04b}. 
 The  $(k,0)$-moment dependence of the 
$m_s$ prediction has been reduced after the recent
OPAL and CLEO analyses finding larger branching fractions
for $\tau^- \to K^- \pi^+ \pi^- \nu$. Accurate experimental data will clarify 
the origin of the remaining moment dependence and will allow a consistency check of 
the whole analysis. 

Another point to be checked~\cite{GJPPS04b} is the fulfilment of 
the quark-hadron duality between the QCD OPE and the OPAL spectral function. 
A first look at this question was presented in ~\cite{Malt04}.

\section*{Acknowledgments}
 This work has been supported in  part by
the European Commission (EC) RTN Network EURIDICE
Grant No. HPRN-CT2002-00311 (A.P. and J.P.),
by German BMBF under Contract No. 05HT4WOA/3 (M.J.),
by MCYT (Spain) and FEDER (EC) Grants No.
FPA2001-3031 (M.J. and A.P.) and FPA2003-09298-C02-01 (J.P.),
  and by Junta de Andaluc\'{\i}a
Grant No. FQM-101 (J.P.). E.G. is indebted
to the EC for a Marie-Curie Grant No. MEIF-CT-2003-501309.


\section*{References}
\begin{thebibliography}{99}

\bibitem{BNP92} E. Braaten, S. Narison and A. Pich, 
Nucl. Phys. {\bf B 373} (1992) 581;
S. Narison and A. Pich, Phys. Lett. {\bf B 211} (1988) 183;
E. Braaten, Phys. Rev. Lett. {\bf 60} (1988) 1606;
Phys. Rev. {\bf D 39} (1989) 1458.

\bibitem{ALEPH98}  ALEPH Collaboration, R. Barate et al.,
Eur. Phys. J. {\bf C 4} (1998) 409;
M. Davier and C. Yuan, Nucl.\ Phys.
 B (Proc.\ Suppl.)  {\bf 123} (2003) 47.

\bibitem{OPAL99} OPAL Collaboration, K. Ackerstaff et al., 
Eur. Phys. J. {\bf C 7} (1999) 571.

\bibitem{CLEO93} CLEO Collaboration, T. Coan et al., 
Phys. Lett.  {\bf B 356} (1995) 580.

\bibitem{ALEPH99}  ALEPH Collaboration, R. Barate et al.,
Eur. Phys. J. {\bf C 11} (1999) 599.

\bibitem{CLEO03} CLEO Collaboration,
R.A. Briere et al.,  Phys. Rev. Lett. {\bf 90} (2003) 181802.

\bibitem{OPAL04} OPAL Collaboration, 
G. Abbiendi et al., Eur. Phys. J. {\bf C 35} (2004) 237.

\bibitem{CHD97} 
M. Davier, Nucl. Phys. B
(Proc.\ Suppl.)   {\bf 55C} (1997) 395;
S. Chen, Nucl. Phys. B (Proc. Suppl.)
{\bf 64} (1998) 256;  
S. Chen, M. Davier and A. H\"ocker,
Nucl. Phys. B (Proc. Suppl.) {\bf 76} (1999) 369.

\bibitem{PP98} A. Pich and J. Prades, J. High Energy
Phys. {\bf 06} (1998) 013;
Nucl. Phys. B (Proc. Suppl.) {\bf 74} (1999) 309;
J. Prades, Nucl. Phys. B (Proc. Suppl.) {\bf 76} (1999) 341.

\bibitem{CKP98} K.G. Chetyrkin, J.H. K\"uhn and A.A. Pivovarov,
Nucl. Phys. {\bf B 533} (1998) 473.

\bibitem{PP99} A. Pich and J. Prades, J. High Energy Phys.
{\bf 10} (1999) 004;
Nucl. Phys. B (Proc. Suppl.) {\bf 86} (2000) 236.

\bibitem{KM00} J. Kambor and K. Maltman,  Phys. Rev. 
{\bf D 62} (2000) 093023;
Nucl. Phys. {\bf A 680} (2000) 155;
Nucl. Phys. B (Proc. Suppl.) {\bf 98} (2001) 314.

\bibitem{CDGHPP01} S. Chen et al., Eur. Phys. J.
{\bf C 22} (2001) 31;
M. Davier et al.,  Nucl. Phys. B (Proc. Suppl.)
{\bf 98} (2001) 319.

\bibitem{GJPPS03}E. G\'amiz et al., 
J. High Energy Phys. {\bf 01} (2003) 060.

\bibitem{GJPPS04a} E. G\'amiz et al., Phys.\ Rev.\ Lett.\  {\bf 94} (2005) 
011803; 
hep-ph/0411278.

\bibitem{Sew}  W. Marciano and  A. Sirlin, Phys. Rev. Lett.
{\bf 61} (1988) 1815;
E. Braaten and C.S. Li, Phys. Rev. {\bf D 42} (1990) 3888;
J. Erler, Rev. Mex. Fis.  {\bf 50} (2004) 200.

\bibitem{CK93} K.G. Chetyrkin and A. Kwiatkowski, 
Z. Phys. {\bf C 59} (1993) 525; Erratum, hep-ph/9805232.

\bibitem{CHE04} P.~A.~Baikov, K.~G.~Chetyrkin and J.~H.~Kuhn, 
hep-ph/0412350.

\bibitem{GJPPS04b} E. G\'amiz et al., in preparation.

\bibitem{MK02} K. Maltman and J. Kambor, Phys. Rev. 
{\bf D 65} (2002) 074013.

\bibitem{JOP02} M. Jamin, J.A. Oller and A. Pich,
Eur. Phys. J. {\bf C 24} (2002) 237.

\bibitem{NAR99} S. Narison, Phys. Lett. {\bf B 466}
(1999) 345.

\bibitem{HAS04} S. Hashimoto, hep-ph/0411126.

\bibitem{lr:84}
H.~Leutwyler and M.~Roos, Zeit. Phys. \textbf{C25} (1984) 91.

\bibitem{e865:03}
E865 Collaboration, A.~Sher et~al., Phys. Rev. Lett. \textbf{91} (2003) 261802.

\bibitem{ktev:04}
T.~Alexopoulos {\it et al.}  [KTeV Collaboration], 
Phys.\ Rev.\ D {\bf 70} (2004) 092007.

\bibitem{ckm:03}
M.~Battaglia et~al.  (2003), hep-ph/0304132.

\bibitem{JOP04}
M.~Jamin, J.~A.~Oller and A.~Pich,
JHEP {\bf 0402} (2004) 047.

\bibitem{Becirevic04}
D.~Becirevic {\it et al.},
Nucl.\ Phys.\ B {\bf 705} (2005) 339.

\bibitem{BT03}
J.~Bijnens and P.~Talavera,
Nucl.\ Phys.\ B {\bf 669} (2003) 341.

\bibitem{CNP04}
V.~Cirigliano, H.~Neufeld and H.~Pichl,
Eur.\ Phys.\ J.\ C {\bf 35} (2004) 53.

\bibitem{Marciano}
W.~J.~Marciano,
Phys.\ Rev.\ Lett.\  {\bf 93} (2004) 231803.

\bibitem{MILC04} 
C.~Aubin {\it et al.}  [MILC Collaboration],
Phys.\ Rev.\ D {\bf 70} (2004) 114501.


\bibitem{CMS04} A. Czarnecki, W.J. Marciano and A. Sirlin,
hep-ph/0406324.

\bibitem{ROS04} J.L. Rosner, hep-ph/0410281.

\bibitem{Malt04}
K.~Maltman, hep-ph/0412326.

\end{thebibliography}
\end{document}